\documentclass[preprint,showpacs,preprintnumbers,amsmath,amssymb]{revtex4}
\usepackage{hyperref}
\usepackage[mathscr]{eucal}
\usepackage{graphicx,wrapfig}

\begin{document}

\title{  Quasi exactly solvable extension of Calogero model associated with exceptional orthogonal polynomials}
\author{B. Basu-Mallick}\email{bireswar.basumallick@saha.ac.in }
\affiliation{ Saha Institute of Nuclear Physics, HBNI,1/AF Bidhannagar, Kolkata-700064, India }
\author{Bhabani Prasad Mandal} \email{bhabani.mandal@gmail.com} 
\affiliation { Department of Physics, 
Banaras Hindu University, 
Varanasi-221005, India} 
\author{Pinaki Roy}\email{pinaki@isical.ac.in}
\affiliation { Physics and Applied Mathematics Unit, Indian Statistical Institute, Kolkata-700108, India.  }

\begin{abstract}
By using the technique of supersymmetric quantum mechanics,
we study a quasi exactly solvable  extension of the $N$-particle
rational Calogero model with harmonic confining interaction. 
Such quasi exactly solvable many particle system, whose effective
potential in the radial direction yields a supersymmetric partner of  
the radial harmonic oscillator,
 is constructed by including new long-range interactions to the rational Calogero model. 
An infinite number of bound state energy
levels are obtained for this system under certain conditions. We also calculate the corresponding bound state wave functions in terms of the recently discovered exceptional orthogonal Laguerre polynomials.

\end{abstract}

\maketitle
%\section{Introduction}

The subject of exactly solvable  many particle quantum systems \cite{ca,ca1,su} with long-range interaction continues to attract much attention due to its connection to diverse subjects like generalized exclusion statistics \cite{ha,is,po}, random matrix theory \cite{tani}, spin-chain models \cite{hal,shas, poly,canew,fink,bbm, bbm1}, Tomonaga-Luttinger liquid \cite{lutt}, quantum chaos \cite{chaos} and black holes \cite{black_hole} and have been extended in various directions \cite{olsh, corr, bk,khare, pkp,ptcal,ptcal1,frin}. 
Rational Calogero model \cite{ca}
which describes $N$ particles on a line with pairwise long-range interacting potential is one of the most important and well known among these exactly solvable many particle quantum systems. Inspired by its successes and wide applicability in various branches of physics, it is of considerable interest to find new exactly solvable $N$-body problem. 

The purpose of the present work is to extend $N$-particle Calogero model by including 
quasi exactly solvable (QES) long-range interactions.  As a tool we shall use the technique of supersymmetric quantum mechanics~\cite{susy}. The many particle Calogero like models have been studied extensively in the framework of supersymmetric quantum mechanice \cite{susycal,2,3,4,susycal3}.
In these works, the total Hamiltonian of the $N$ particle
Calogero like  models have been expressed through $N$ pairs of creation
and annihilation operators. The supersymmetric partner Hamiltonians obtained in this
way exhibit the shape invariant property. As a result, the full spectra of
such Calogero like models are derived by using supersymmetric quantum mechanics.
The Lax operators of such Calogero like models and their conserved
quantities have also been constructed by using this approach.

The QES potential, which we consider in this paper
along with the inverse square interaction term of the original Calogero
model, depends only on the radial coordinate $r$. So the angular part of
the corresponding wave function can be solved exactly like the case of
original Calogero model \cite{ca,ca1}. Therefore, only the radial part of the wave
function is important in this case and our aim is to derive
such radial wave functions by using the method of supersymmetric quantum mechanics.  We follow the original work of Calogero
to split the Hamiltonian of the $N$ particle system into the radial part
and angular parts. Subsequently, we apply the techniques of the supersymmetric
quantum mechanics to the radial effective Hamiltonian by using only one
pair of creation and annihilation operators to construct a QES $N$-particle system, whose effective
potential in the radial direction would become a supersymmetric partner to the exactly solvable radial harmonic oscillator potential and represent a conditionally
exactly solvable system  where two parameters are related in a
certain way. The supersymmetric  quantum
mechanics is applied only for finding out the spectrum of the radial
effective Hamiltonian. Furthermore, there is no shape invariance among the
supersymmetric partner Hamiltonians in this case.  This fact plays an
important role in our construction of a $N$-particle QES Hamiltonian with
long-range interaction.
Due to supersymmetry, exact expressions for an  
infinite number of bound state energy
levels of this QES many particle system with conditionally exactly solvable effective potential
can be derived analytically. Indeed, within a certain sector, such QES many particle system would become iso-spectral to the Calogero model for which exactly solvable radial harmonic oscillator acts 
as the effective potential. We shall also explicitly calculate the wave functions for this QES system and it will be shown that the wavefunctions can be expressed in terms of the recently discovered exceptional Laguerre orthogonal polynomials \cite{eop1,eop2,eop3,rr,eop_bpm,sasaki,sasaki1,ho,tanaka, quesne,bagchi, granditi,ces1}.

We start with a Calogero like many particle system with arbitrary potential $U(\sqrt{N} r)$ depending on the `radial' coordinate  $r$, described by the Hamiltonian,

\begin{align}
H=-\sum_{i=1}^N \frac{\partial^2}{\partial x_i^2} +\sum_{i<j}^N\frac{g}{(x_i-x_j)^2}+ U(\sqrt{N} r) 
\label{hc}
\end{align}
where $g\geq -\frac{1}{2}$ and the radial coordinate is defined as
$r=\sqrt{ \frac{1}{N}\sum_{i<j}^N (x_i-x_j)^2}$. Our aim is to 
find out some suitable form of $U(\sqrt{N} r)$, which would lead to 
exactly solvable normalizable  solutions of the eigenvalue problem 
\begin{equation}
H\psi(x) =E\psi(x) \, , 
\label{eigen}
\end{equation}
where $x\equiv (x_1,x_2, \cdots, x_N)$. To this end, 
we follow the standard method \cite{ca} and 
make an ansatz for $\psi(x)$ in a sector of configuration space corresponding
to a definite ordering of particles (e.g., $x_1\geq x_2 \geq \cdots
\geq x_N$) as
\begin{equation}
\psi(x)= \prod_{i<j}(x_i-x_j)^{a+\frac{1}{2}} P_{k,q}(x)\phi(r),     
\label{ansa}
\end{equation}
where $a=\frac{1}{2}\sqrt{1+2g}$ and 
$P_{k,q}(x)$'s are translationally invariant, symmetric, $k$-th order
homogeneous polynomials satisfying the differential equations 
\begin{equation}
\sum_{j=1}^N \frac{\partial^2 P_{k,q}(x)}{\partial x_j^2} +
\left(a+\frac{1}{2}\right)\sum_{j \neq
k}\frac{1}{(x_j-x_k)}\left(\frac{\partial}{\partial x_j} -
\frac{\partial} {\partial x_k}\right) P_{k,q}(x) = 0\, .
\label{b6} 
\end{equation}
Note that the index $q$ in $P_{k,q}(x)$ can take
any integral value ranging from $1$ to $g(N,k)$, where $g(N,k)$ is
the number of independent polynomials.
The existence of such translationally invariant, symmetric and 
homogeneous polynomial solutions of (\ref{b6}) has been
discussed by Calogero \cite{ca}.
The radial part of the wave function, $\phi(r)$ then satisfies the differential equation
\begin{equation}
-[\phi^{\prime\prime }(r)+2(k+b+1)\frac{1}{r}\phi^\prime (r)] +[U(\sqrt{N}r)-E]\phi(r)=0
\label{phi}
\end{equation}
with $b=\frac{N(N-1)}{2}a +\frac{N(N+1)}{4}-2$.  
To solve this radial equation we further substitute

\begin{equation}
\phi(r)= r^{-(l+1)}\chi(r), \ \ \  l=k+b
\label{kai}
\end{equation}
in Eq.(\ref{phi}) to obtain

\begin{equation}
-\chi^{\prime\prime}(r) +
\left[ \frac{l(l+1)}{r^2} + U(\sqrt{N}r)\right]\chi (r) = E\chi (r) \, .
\label{chi}
\end{equation}
Thus, for any given value of $k$,  the eigenvalue problem of the many particle Hamiltonian 
(\ref{hc}) is reduced to that of an effective
single particle Hamiltonian of the form 
\begin{equation}
H_{eff}= -\frac{d^2}{dr^2}+U_k(\sqrt{N}r),
\label{eff}
\end{equation}
 where   
\begin{equation}
 U_k(\sqrt{N}r)=\frac{l(l+1)}{r^2} + U(\sqrt{N}r) \, .
 \label{uk}
\end{equation}
Combining Eqs.~(\ref{ansa}) and (\ref{kai}), we can express the 
eigenfunctions of $H$ in (\ref{hc}) through those of $H_{eff}$ in (\ref{eff}) as 
\begin{equation}
\psi(x)= r^{-(l+1)}\prod_{i<j}(x_i-x_j)^{a+\frac{1}{2}} P_{k,q}(x)\chi(r),     
\label{ansa1}
\end{equation}

Next, we will be using supersymmetric techniques to construct a novel form of $ U(\sqrt{N}r)$,  for which a class of eigenfunctions of 
the many particle Calogero like Hamiltonian (\ref{hc}) can be obtained exactly.
%under certain condition. 
To this end we recall the supersymmetric formulation of quantum mechanics \cite{susy}, according to which a pair of Hamiltonians 

\begin{equation}
H_{\pm}= A^{\pm}A^{\mp} = \left[-\frac{d^2}{dr^2} + V_{\pm}(r)\right] \ \mbox{  with }  \ \  V_{\pm}(r)= W^2(r)\pm W^\prime(r)
\label{susy1}
\end{equation}
where $W(r)$ is called the superpotential and
\begin{equation}
A^\pm = \left(\pm \frac{d}{dr}+W(r)\right)
\label{aa}
\end{equation}
forms a supersymmetric system. The above Hamiltonians are iso-spectral to each other except 
the ground state energy (normalized to be zero)  associated to $H_-$ for unbroken SUSY. The relationship between the energies and corresponding eigenfunctions of these Hamiltonians are given by
\begin{subequations}
\label{susy}
\begin{align}
\mbox {Unbroken Supersymmetry : \ \ \ \ \ \ \ \ } \  \ \ \ \ \ \ \ E_0^-=0\ ; \ \ \ \ E_{n+1}^-=E_n ^+>0    \\  \nonumber
\psi_0=N e^{-\int W(r)dr} ,\ \  \psi_n^+=\frac{1}{\sqrt{E^-_{n+1}}}A^+\psi_{n+1}, \ \  \psi_{n+1}^-=\frac{1}{\sqrt{E^+_{n}}}A^-\psi_{n}\label{ubsusy} \, .
%\end{align}
\\ 
%\begin{align}
\mbox{Broken Supersymmetry:                                                    }  \ \ \ \ \ \ \ \ \ \ \ \ \ \ \ \ \ \ \ \ \ \ \ \ \ \ \ \ \ \  \ \ \ \ \ \ \ \ \ \ \ \ \ \ \ \ \ \ \ \ \ \ \ \ 
\nonumber \\
E_{n}^-=E_n ^+>0  , \ \  \psi_n^+=\frac{1}{\sqrt{E^-_{n}}}A^+\psi_{n}, \ \  \psi_{n}^-=\frac{1}{\sqrt{E^+_{n}}}A^-\psi_{n} \, .
\end{align}
\label{bsusy}
\end{subequations}

Now  let us consider a specific superpotential of the form \cite{ces}
\begin{equation}
W(r)= r+\frac{2g_1 r}{1+g_1 r^2} + \frac{\alpha+1}{r} \, ,
\label{w1}
\end{equation} 
where  $\alpha$ is a real positive parameter and 
the other parameter $g_1$ depends on
$\alpha $ as 
\begin{equation}
g_1=\frac{2}{2\alpha+3} \, .
\label{ces}
\end{equation}
It is easy to see that for the above choice of superpotential, supersymmetry is broken. Now using Eq.(\ref{susy1}) the partner potentials can be found to be
\begin{subequations}
\label{vv}
\begin{align}
&V_+(r)= r^2 +\frac{\alpha(\alpha+1)}{r^2}+2\alpha+7 \, , 
\label{v+}
\\
&V_-(r) = r^2 +\frac{(\alpha+1)(\alpha+2)}{r^2}-\frac{ 4g_1}{1+g_1 r^2} +\frac{8g_1^2 r^2}{(1+g_1 r^2)^2} + 2\alpha +5 \, . 
\label{v-}
\end{align}
\end{subequations}

One of the above mentioned partner potentials, namely, $V_+(r)$ 
is nothing but radial harmonic oscillator and the complete solution of the corresponding eigenvalue equation is given as,
\begin{equation}
E_n^+= 4\left(n+\alpha +\frac{5}{2}\right),~~~~n=0,1,2,\cdots 
\end{equation}
\begin{equation}
\chi_n^+(r) = \sqrt{\frac{n!}{\Gamma(n+\alpha+\frac{3}{2})}} \,  r^{\alpha+1} e^{-r^2/2} L^{\alpha+1/2}_n(r^2) \, .
\end{equation} 
Now using the results of Eq.(\ref{bsusy}) for broken supersymmetry, we construct the solution for the partner system with potential $V_-(r)$ given in Eq.(\ref{v-}) as
\begin{equation}
E_n^-= 4\left(n+\alpha +\frac{5}{2}\right) \, ,
\label{eigenv}
\end{equation}
\begin{equation}\label{wf}
\chi_n^-(r) = \sqrt{\frac{1}{E_n^+} }A^- \chi_n^+(r) \, .
\end{equation} 
 Then using Eqs.(\ref{aa}) and (\ref{w1}) the wavefunction in Eq.(\ref{wf}) can be written as
\begin{eqnarray}
\chi_n^-(r) &&= \sqrt{\frac{4(n!)}{E_n^+\Gamma(n+\alpha+\frac{3}{2})}} r^{\alpha+2} e^{-r^2/2} \left[\frac{1+g_1+g_1r^2}{1+g_1r^2}L_n^{\alpha+1/2} (r^2)+L_{n-1}^{\alpha+3/2}(r^2)\right] , \nonumber \\ && (\mbox {with }  L^\alpha_{-1}=0)
\end{eqnarray}
where we have used the property $ \frac{d^k}{dr^k} L_n^\alpha (r) =(-)^k L^{\alpha+k}_{n-k}(r)\  \mbox{ if } k\le n,\ \mbox{ and } =0\ \mbox{ for } \ k>n$ for usual Laguerre polynomials.
Now using the relation \cite{eop3,rr}
\begin{equation}
\hat{L}^k_{n,m}(r^2)= L_m^k(-r^2)L^{k-1}_{n-m}(r^2) + L_m^{k-1}(-r^2)L^{k}_{n-m-1}(r^2),  \ \ \  n\ge m
\label{rela} 
\end{equation}
where $\hat{L}^k_{n,m}(r^2)$ denotes the Exceptional $X_m$ Laguerre polynomial, the above wavefunction can be written in terms of Exceptional Laguerre polynomial
as
\begin{equation}
\chi_n^-(r) = \sqrt{\frac{4(n!)}{E_n^+\Gamma(n+\alpha+\frac{3}{2})}} r^{\alpha+2} e^{-r^2/2} 
\frac{1}{L_1^{\alpha+1/2}(-r^2) } \hat{L}^{\alpha+3/2}_{n+1,1}(r^2).
\label{s=}
\end{equation} 
However, it should be stressed that, the eigenvalue problem 
associated with the partner
potential $V_-(r)$ in (\ref{v-}) can be solved exactly 
in the above mentioned way only when the parameter $g_1$ 
depends on $\alpha$ through the relation (\ref{ces}). Hence 
the Hamiltonian corresponding to $V_-(r)$ represents a conditionally exactly solvabl \cite{ces1}. 

It is well known that $V_+(r)$ in (\ref{v+}) 
 can be used to generate the
 exactly solvable Calogero model with harmonic confining interaction. 
Indeed, equating the effective
 potential $U_k(\sqrt{N}r)$
 in (\ref{uk}) with the radial harmonic oscillator
 potential $V_+(r)$ in (\ref{v+}) 
 (up to some additive constant), one obtains 
 \begin{equation}
  U(\sqrt{N}r)= r^2 + \frac{\alpha(\alpha+1)-l(l+1)}{r^2} ,
  \label{calp}
 \end{equation}
 where $l=k+b$ and $\alpha$ is a free parameter. Even though $U(\sqrt{N}r)$ in (\ref{calp}) depends
 on $k$ for an arbitrary choice of the parameter $\alpha $, 
 one can take    
 $\alpha=l$ and obtain  $ U(\sqrt{N}r)= r^2 $ which no longer depends 
 on $k$. By essentially using this procedure 
 and taking advantage of the fact that $V_+(r)$ in (\ref{v+}) represents
 an exactly solvable system, 
 Calogero has shown that the many particle 
 system (\ref{hc}) with $U(\sqrt{N}r)=r^2$  
 can be solved exactly for all possible values 
 of $k$. 
 
 For our present purpose, we equate the effective
 potential $U_k(\sqrt{N}r)$
 in (\ref{uk}) with the conditional exactly solvable potential   
 $V_-(r)$ in (\ref{v-}), i.e., we take  
 $U_k(\sqrt{N}r)=V_-(r) $. In this way, we obtain a new form of 
$ U(\sqrt{N}r)$ given by 
 \begin{equation}
   U(\sqrt{N}r)= r^2 +\frac{(\alpha+1)(\alpha+2)-l(l+1)}{r^2}-\frac{ 4g_1}{1+g_1 r^2} +\frac{8g_1^2 r^2}{(1+g_1 r^2)^2} + 2\alpha +5 \, ,
   \label{calp1}
 \end{equation}
 where $l=k+b$, $\alpha$ is a free parameter and  $g_1$ 
depends on $\alpha$ through the relation (\ref{ces}).
Since $g_1$ depends on $\alpha$, it is clearly not possible to choose
any special value of the parameter $\alpha$ 
(like, for example, $\alpha=l-1$) so that 
%in contrast to $U(\sqrt{N}r)$ in (\ref{calp}), 
 $U(\sqrt{N}r)$ in (\ref{calp1}) becomes independent of $k$.  
 Hence, the many particle 
 system (\ref{hc}) with $U(\sqrt{N}r)$ given in  (\ref{calp1}) 
 represents a QES, which can be solved exactly only for a particular
 value of $k$.  From the above discussion it is evident that, 
the conditionally exactly solvable nature of single particle 
potential $V_-(r)$ in (\ref{v-}) is responsible for the QES nature
of the many particle 
 system (\ref{hc}) with $U(\sqrt{N}r)$ given in  (\ref{calp1}). 
Using the known eigenvalues and eigenfunctions of the
single particle system associated with $V_-(r)$,
we shall now derive infinite number of exact eigenvalues
 and eigenfunctions (related to 
 a particular value of $k$) for the many particle 
 system (\ref{hc}) with $U(\sqrt{N}r)$ of the form  (\ref{calp1}).
 Indeed, by using Eq.~(\ref{eigenv}), it is easy to obtain such 
 eigenvalues as 
\begin{equation}
E_n= 4\left(n+\alpha +\frac{5}{2}\right),~~~~n=0,1,2,\cdots . 
\label{eigenv2}
\end{equation}
Furthermore, substituting the expression of $\chi_n^-(r)$ given in 
Eq.~(\ref{s=}) to the place of $\chi(r)$ in Eq.~(\ref{ansa1}), 
we obtain the corresponding (unnormalized) eigenfunctions as 
 \begin{equation}
\psi_n(x)= r^{\alpha -l+1}
e^{-r^2/2} 
\frac{1}{L_1^{\alpha+1/2}(-r^2) } \hat{L}^{\alpha+3/2}_{n+1,1}(r^2)
\prod_{i<j}(x_i-x_j)^{a+\frac{1}{2}} P_{k,q}(x) \, . 
\label{eigenf2}
\end{equation}
Since the ground state of the original Calogero model is obtained 
in the special case $k=0$ and $n=0$, it is interesting
to check the form of eigenvalue (\ref{eigenv2}) and the corresponding eigenfunction 
(\ref{eigenf2}) for this special case. By putting $n=0$ in (\ref{eigenv2}), we obtain 
  $E_0= 4\left(\alpha +\frac{5}{2}\right)$.
  Moreover, by putting $k=0$ and $n=0$ in (\ref{eigenf2}),
  and also using the fact that $l=b$ and  $P_{k,q}(x)=1$ for $k=0$, 
  we obtain the corresponding eigenfunction as 
\[
\psi_0(x)= r^{\alpha -b+1}
e^{-r^2/2} 
\frac{1}{L_1^{\alpha+1/2}(-r^2) } \hat{L}^{\alpha+3/2}_{1,1}(r^2)
\prod_{i<j}(x_i-x_j)^{a+\frac{1}{2}}  \, . 
\label{grst}
\]
Using Eq.~(\ref{rela}), one can further simplify the above 
eigenfunction as 
\begin{equation}
\psi_0(x)= r^{\alpha -b+1}
e^{-r^2/2} 
\left(\frac{r^2+\alpha+5/2}{r^2+\alpha+3/2} \right) 
\prod_{i<j}(x_i-x_j)^{a+\frac{1}{2}}  \, . 
\label{grst}
\end{equation}

  It is interesting to note that our results
  for the single particle conditionally exactly solvable potential $V_-(r)$ in Eq.~(\ref{v-})
  are consistent with  the results obtained through point canonical transformation  approach\ cite{pct} to extend the radial
  harmonic oscillator rationally.  Using this approach one can obtain the most general rationally extended radial oscillator potential \cite{rr}
  \begin{eqnarray}
  V_m(r) = r^2 +\frac{l(l+1)}{r^2}- \frac{4r^2 L^{l+3/2}_{m-2}(-r^2)}{L_m^{l-1/2}(-r^2)}+2(2r^2 +2l-1)\frac{4r^2 L^{l+1/2}_{m-2}(-r^2)}{L_m^{l-1/2}(-r^2) } \nonumber \\
  +8 r^2 [\frac{4r^2 L^{l+3/2}_{m-2}(-r^2)}{L_m^{l-1/2}(-r^2)}]^2-4  
  \label{vm}
  \end{eqnarray}
  whose bound state solutions for the energy eigenvalues $E_n = ( 4n+2l+3)\  \  n=0,1,2\cdots $ are written in terms of $X_m$ exceptional Laguerre Polynomials as
  
  \begin{equation}
  \chi_{n,m}(r^2)= N_{n,m}\frac{x^{l+1}\exp(-r^2/2)}{L_m^{l-1/2}(-r^2)}\hat{L}^{l+1/2}_{n+m,m} (r^2)\ \ 
  \label{psipct}  
  \end{equation}
  
  Where
  $
  N_{n,m}= [\frac{(n-m)!}{(l+1/2+n)\Gamma(l+1/2+n-m)}]^{1/2} $ and $ \hat{L}^{l+1/2}_{n+m,m} (r^2)$ is    $X_m$  exceptional orthogonal Laguerre polynomial, $m=0$ corresponds to usual Laguerre polynomials.
  Now we note that $m=1$ corresponds to the case we discus in the context of Calogero Model as
  the potential 
  \begin{equation}
  V_1= r^2 +\frac{l(l+1)}{r^2} -\frac{8}{2r^2 +2l +1} + \frac{32r^2}{(2r^2+2l +1)^2}
  \end{equation}
  calculated from Eq.(\ref{vm}) is same as our $V_-(r)$ in 
  Eq.~(\ref{v-}) when $l=(\alpha+1)$
  and $g_1=\frac{2}{(2\alpha+3) }$ after from a over constant term.
  The solution obtained in point canonical transformation approach in Eq.(\ref{psipct}) for $m=1$ is exactly same as we obtained in 
  supersymmetric approach in Eq.~(\ref{s=}).

In this article we have obtained an exactly solvable $N$ body problem with non trivial pairwise interacting potential. The potential is actually quasi exactly solvable since we have obtained solutions for particular values of the parameter $k$ and at the same time it is also conditionally exactly solvable since the parameter $g_1$ can assume only a particular value. In this context we would like to note that one may consider more general superpotentials of the form \cite{ces1,ces, ces2}
\begin{equation}
W_m(r)=r+\frac{\alpha+1}{r}+\sum_{i=1}^m \frac{2g_ir}{1+g_ir^2} \, , 
\label{superp} 
\end{equation}
where $m$ is any positive integer, and derive a class of conditionally exactly solvable 
single particle potentials  by using the methods of supersymmetry formalism. Similar to the $m=1$ case which has been considered in this paper, the solutions of such conditional exactly solvable potential with any value of $m$ can also be expressed in terms of the recently discovered exceptional orthogonal polynomials of the Laguerre type. It is worth while to 
explore whether, for an arbitrary value of $m$,  the  conditionally exactly solvable
single particle potential obtained from the superpotential
(\ref{superp}) match with  $ V_m(r)$ in (\ref{vm})
which is obtained through the point canonical transformation approach. Furthermore, 
it would be interesting to use such conditionally exactly solvable single particle potentials
to construct non trivial QES many body potentials by following 
the approach of this paper.  Finally it may be noted that in this 
paper we have used a superpotential with broken supersymmetry to 
construct a QES Calogero like many particle system.  
We feel it would be of interest to investigate similar many 
particle systems for which supersymmetry is unbroken.

{\bf Acknowledgements:} One of us (BPM) acknowledges support from theory division, SINP, Kolkata for a visitation, during which a part of the work has been carried out.
%%%%
%%%%%%

\end{document}